\documentstyle[12pt]{article}

\def\ket#1{| #1 \rangle}
\def\bra#1{\langle #1 |}
\def\BraKet#1#2{\langle #1 | #2 \rangle}
\def\KetBra#1#2{| #1 \rangle\langle #2 |}
\def\Tr{{\rm Tr}}
\def\ve{\varepsilon}
\def\Z{{\cal Z}}
\def\z{{ z}}

\begin{document}
\centerline{\large\bf On the teleportation of continuous variable}
\vskip 3mm
\centerline{S.N.Molotkov and S.S.Nazin}
\vskip 2mm
\centerline{\sl\small Institute of Solid State Physics 
            of Russian Academy of Sciences}
\centerline{\sl\small Chernogolovka, Moscow district, 142432,  Russia}
\vskip 3mm
\begin{abstract}
The measurement procedures used in quantum teleportation
are analyzed from the viewpoint of the general theory of 
quantum-mechanical measurements.
It is shown that to find the teleported state one should only 
know the identity resolution (positive operator-valued measure) 
generated by the  corresponding instrument (quantum operation 
describing the system state change caused by the measurement)
rather than the instrument itself.
A quantum teleportation protocol based on a measurement 
associated with a non-orthogonal identity resolution is proposed
for a system with non-degenerate continuous spectrum. 
\end{abstract}

\vskip 3mm

\section{Introduction}
One of the major results of quantum information theory consists in
the possibility of teleportation of an unknown quantum state
by means of a classical and a distributed quantum communication channel, 
the latter being realized by a non-local entangled state chosen in a 
special way (e.g., an EPR-pair of particles) \cite{Bennett93}. 
Quantum teleportation of an unknown state from user A to user B
is performed in the following way: \cite{Bennett93}. User A has
the unknown state $\rho_1$ of quantum system 1 
(e.g., spin-1/2 particle; Ref.\cite{Bennett93} contains also
the more general case of a quantum system with arbitrary finite
number of levels, i.e. with any finite-dimensional state space)
which is to be teleported to user B. In addition, two other
(also spin-1/2) particles labeled as systems 2 and 3 are employed 
which are in the spin-entangled EPR-state $\rho_{23}$ such that
the user A has access to particle 2 while user B has access to particle 3.
User A performs a certain joint measurement $m_{12}$ over the
system 1 in the unknown state $\rho_1$ and particle 2 from the EPR-pair.
As a result, the total system composed of particles 1, 2, and 3
changes its state from $\rho_1\otimes\rho_{23}$  to a new state
$\rho'_{123}$ which depends on the measurement result $\z$.
It turns out that there exist such measurements $m_{12}$ that
the state $\rho'_3$ of particle 3 from the EPR-pair accessible to user B 
after the measurement (which is obtained from $\rho'_{123}$ by
performing trace of the state spaces of particles 2 and 3,
$\rho'_3=\mbox{Tr}_{1,2}\rho'_{123}$) is related to the initial state
$\rho_1$ of particle 1 through a certain unitary transformation
$U_\z$ which does not depend on $\rho_1$ and is completely determined
by the result $\z$ of the performed measurement $m_{12}$:
\begin{equation}
  \rho'_3=U_\z\rho_1 
\end{equation} 
(here and later we identify the isomorphic state spaces of particles 1 and 3).
The classical communication channel is necessary for user A to
convey to user B the measurement result $\z$ which tells him which
unitary transformation $U^{-1}_\z$ should be applied to the state $\rho'_3$
of particle 3 to recover the state $\rho_1$.
Note that the user A does not obtain any information on the teleported state.

The outlined algorithm of quantum teleportation substantially employs the
fact that after the measurement is performed, the system as a whole 
(all three particles) is described by a certain state $\rho'_{123}$
determined by the measurement result; the algorithm of Ref.\cite{Bennett93} 
uses the so-called Bell measurement described by a certain self-adjoint
operator with non-degenerate spectrum in the 4-dimensional state space
(of particles 1 and 2) and the state 
$\rho'_{123}$ can be easily written explicitly. 

The first algorithm for teleportation of quantum continuous variable
(i.e. the wave function of a one-dimensional non-relativistic
spinless particle whose state space is infinite-dimensional)
was given by Vaidman \cite{Vaidman94}. Later this approach was extended
to a more realistic algorithm of the teleportation of a single-mode
electromagnetic field \cite{BraunKimble98}. Both these algorithms 
actually assumed that in the case of an observable with a continuous spectrum
the state of the system just after the measurement is described 
by the ``eigenvector'' belonging to the ``eigenvalue'' (of the
corresponding self-adjoint operator) given by the measurement act.

However, for a continuous variable the correctly formulated question 
concerning the system state after the measurement turns out to be much more 
difficult than in the case of a discrete spectrum (e.g. see \cite{Ozawa93}).
The problem here is not even only that for a continuous spectrum the Hilbert
state space does not contain any correctly defined eigenvectors.
Consider, for example, a self-adjoint operator $A$ with the continuous 
spectrum $\Lambda$. Let the point $\z$ belong to this spectrum
and the system state before the measurement be $\rho$.
How sensible is then the question of what is the system state $\rho_\z$ after 
the measurement which gave the result $r=\z$?
The problem is that according to the statistical interpretation
of quantum mechanics the very concept of ``state'' should only be
associated with an ensemble of identical systems rather than single systems.
In our case it is then natural to think that one should consider the 
subensemble of systems selected by the condition $r=\z$ after the measurement.
However, for a continuous spectrum the probability of obtaining any particular
result $\z$ is zero since any point has zero measure. Therefore, it is simply 
impossible to select the subensemble of systems by the condition $r=\z$,
since the probability of obtaining the same results in any two measurements
is zero. Therefore, the problem of interpretation of the physical meaning 
ascribable to $\rho_\z$ is not so straightforward. To analyze this problem
we shall need some concepts of the general theory of quantum-mechanical 
measurements (e.g. see \cite{Ozawa93,Kraus,Holevo}). Basic ideas and
some results of this theory are outlined in Section 2. In Section 3
the general theory is applied to a particular class of measurements used 
in quantum teleportation. Section 4 considers the teleportation protocol
for a continuous variable presented in Ref.\cite{Vaidman94} within the
framework of the results obtained in Section 3. Section 5 contains a
new algorithm for the teleportation of the states of a model
system with a continuous spectrum based on a measurement associated 
with a non-orthogonal identity resolution. Finally, the last Section 6
summarizes the results obtained in the paper.

\section{Quantum-mechanical measurements}
For a quantum system with a finite-dimensional Hilbert state space ${\cal H}$ 
(when any operator has a purely discrete spectrum), the canonical (von Neumann)
measurement of the observable corresponding to a self-adjoint operator $A$
whose eigenvalues are $\lambda_i$, $i=1\ldots n$ results in the transformation
of the system state from $\rho$ (density matrix just before the measurement) 
to $\rho_j$ if the measurement result is $\lambda_j$
(L\"uders-von Neumann reduction postulate \cite{Neumann,Luders}): 
\begin{equation}
\label{CanonMap}
  \rho \rightarrow \rho_j= 
  \frac{E_j\rho E_j}{\Tr\{E_j\rho\}}. 
\end{equation}
Here $E_j$ is the orthogonal projector on the subspace associated with
the eigenvalue $\lambda_j$ so that the following identity resolution 
takes place:
\begin{equation}
  \label{CanonResolution}
  \sum_j E_j=I,
\end{equation}
where $I$ is the identity operator on $\cal H$, 
and the spectral representation of the operator $A$ is
\begin{equation}
\label{CanonRepresentation}
  A=\sum_j \lambda_j E_j.
\end{equation}
The probability of obtaining $j$-th results is
\begin{equation}
\mbox{Prob}(\lambda_j)=\mbox{Tr}\{\rho E_j\}=\mbox{Tr}\{E_j \rho E_j\}.
\end{equation}

Consider now the most general case when the complete set of all
possible measurement results constitute a measurable space $\Z$
with measure $d\z$ and the quantum system $S$ is described by the
(generally speaking, infinite-dimensional) Hilbert space $\cal H$,
i.e. the set of all its states can be identified with the set $K({\cal H})$ 
of all positive operators on $\cal H$ with trace 1 (i.e., density matrices;
an operator $A$ on $\cal H$ is called positive if $\bra{v} A \ket{v} \ge 0$
for all $v \in  \cal H$).
The set $K({\cal H})$ is a subset of the space
$B_1({\cal H})$ of all finite trace operators on $\cal H$.
In that case the adequate mathematical entity completely
characterizing any particular measurement procedure with the
space $\Z$ of all possible results which can be applied to 
$S$ is the {\sl instrument} \cite{Ozawa93} (or, in a different
terminology, {\sl operation} \cite{Kraus})
$\bf T$, which is a map $\Delta\rightarrow{\bf T}(\Delta)$
of the set $\Gamma$ of all measurable (with respect to the measure $d\z$) 
subsets $\Delta \subset \Z$ into the set of all trace decreasing
(more strictly, non-increasing)
completely positive operators
$P(B_1({\cal H}))$  which map $B_1({\cal H})$ into itself
and satisfies the following two requirements:\\
1) ${\bf T}(\Delta)=\sum_j {\bf T}(\Delta_j)$, if $\Delta=\cup_j \Delta_j$, 
$\Delta_j\cap \Delta_i=\emptyset$ for $i \neq j$ (additivity) and\\ 
2) ${\rm Tr} \{{\bf T}(\Z)\rho\}={\rm Tr} \rho$ for any
$\rho \in B_1({\cal H})$ (normalization).\\
Remember that the linear map $\bf F$ from $B_1({\cal H})$ into
itself is called completely positive if ${\bf F}(L)>0$ for any
$L>0$ from $B_1({\cal H})$, i.e. if it maps any positive operator
from $B_1({\cal H})$ into a positive operator, and in addition, 
possesses a property that if ${\cal H}_0$ is another Hilbert 
space then the map ${\bf F}\otimes{\bf I}$:
$B_1({\cal H}\otimes{\cal H}_0)\rightarrow B_1({\cal H}\otimes{\cal H}_0)$,
defined on the elements of the type $W\otimes W_0 \in B_1({\cal H}\otimes
{\cal H}_0)$ by a formula 
${\bf F}\otimes{\bf I}\, (W\otimes W_0) = {\bf F}(W)\otimes W_0$ and extended 
to the entire space
$B_1({\cal H}\otimes{\cal H}_0)$ by the linearity,
where ${\bf I}$ is the identity operator on $B_1({\cal H}_0)$,
is also a positive map for any ${\cal H}_0$. 
The essence of the instrument {\bf T} is that for any measurable 
subset $\Delta \subset \Z$, $\Delta \in \Gamma$ the state 
$\rho_{\Delta}$ of the subensemble of the systems
initially prepared in the state $\rho \in K({\cal H})$ and then
selected after a repeated application of the specified measurement procedure
by the condition that the measurement result $r=\z$ belongs to the set $\Delta$, 
is (for brevity we write ${\bf T}(\Delta)\rho$ instead of $[{\bf T}(\Delta)](\rho)$)
\begin{equation}
\label{StateAfter}
   \rho_{\Delta} =  \frac{\tilde{\rho}(\Delta)}
                         {\Tr\{\tilde{\rho}(\Delta)\}}=
                    \frac{{\bf T}(\Delta)\rho}
                         {\Tr\{{\bf T}(\Delta)\rho\}} 
                    \in K({\cal H}),\quad
                    \tilde{\rho}(\Delta) = {\bf T}(\Delta)\rho
\end{equation}
while the probability of obtaining result $r = \z \in \Delta$ is 
\begin{equation}
\label{Prob_T}
  {\rm Prob}(\z \in \Delta) = \Tr\{{\bf T}(\Delta)\rho\}
                           = \Tr\{\tilde{\rho}(\Delta)\}; 
\end{equation}
here and later we label by the tilde symbol the ``unnormalized density 
matrices'' (positive operators with trace $\le 1$) which arise after
the application of the operator ${\bf T}(\Delta)$ corresponding to the
considered instrument to the initial density matrix $\rho$.
We shall also apply the term ``density matrices'' to these ``unnormalized
density matrices'' in the cases where it cannot cause confusion.

It is easily checked that for a fixed ${\bf T}$ the formula (\ref{Prob_T})
generates an affine map of the convex set  
$K(\cal H)$ of all possible states $\rho$ of the system $S$ into
the set of probability measures $\nu_{\rm Prob}$ defined on
$\Z$: each state $\rho \in K(\cal H)$ corresponds to the measure
$\mu_{\rho}$ on $\Z$ such that for every set 
$\Delta \in \Gamma$ its measure $\mu_{\rho}(\Delta)$ 
is exactly ${\rm Prob}(\z \in \Delta)$.
It is known \cite{Holevo} that the set of all such maps  
$\rho \rightarrow \mu_\rho$
from $K({\cal H})$ into $\nu_{\rm Prob}$ is in one-to-one correspondence with
the families of Hermitian operators $M(\Delta)$, 
$\Delta \in \Gamma$ defined on the Hilbert state $\cal H$ and satisfying the 
following requirements:\\ 
\noindent
$1'$) $M(\emptyset)=0$, $M(\Z)=I$, (normalization)\\
\noindent
$2'$) $M(\Delta)\ge 0$, (positivity) and \\
\noindent
$3'$) $M(\Delta)=\sum_jM(\Delta_j)$, if $\Delta=\cup_j \Delta_j$, 
$\Delta_j\cap \Delta_i=\emptyset$ for $i \neq j$ (additivity),\\ 
i.e. with the identity resolutions on $\Z$ with the values in the set
of positive operators on ${\cal H}$. The measure $\mu_\rho$ of the set 
$\Delta$ is given by
\begin{equation}
  \label{Prob_M}
  \mu_{\rho}(\Delta) = {\rm Prob}(\z \in \Delta) = \Tr\{\rho M(\Delta)\}.
\end{equation}
In other words, $M(\Delta)$ defines a positive operator-valued measure. 
A special case of the measures of that kind is given by the identity 
resolutions corresponding to the families of spectral projectors
associated with the self-adjoint operators in ${\cal H}$ which, in addition,
possess the property
\begin{displaymath}
M(\Delta_1)M(\Delta_2)=0,  \mbox{\ if\ } \Delta_1\cap \Delta_2=0;
\end{displaymath}
it is natural to call the measurements described by these identity 
resolutions the ``orthogonal measurements''.

Therefore, if we are only interested in the probability distribution of 
obtaining a particular result and do not touch a much more difficult
problem of the system state after the measurement, it is sufficient to 
restrict ourselves to the analysis of the positive identity resolutions
rather than the families of operators ${\bf T}(\Delta) \in P(B_1({\cal H}))$.
The relationship between them is established by the requirement that
the probability of obtaining the result $\z \in \Delta$ after the measurement
act is performed on the system $S$ in any initial state $\rho$ originally 
defined by  Eq. (\ref{Prob_T})  can be calculated with 
the operator $M(\Delta)$ employing Eq. (\ref{Prob_M}).
Comparing Eqs. (\ref{Prob_T}) and (\ref{Prob_M}),
one can easily see that they are compatible if and only if
\begin{equation}
  M(\Delta) = [{\bf T}(\Delta)]^*I, 
\end{equation}
where asterisk means the dual map from the space
$B({\cal H})$ into itself and $I \in B({\cal H})$ is the identity operator
on $\cal H$ (remember that the linear space of all 
bounded operators $B({\cal H})$ on $\cal H$ is isomorphic
to the dual space of $B_1({\cal H})$, and the corresponding isomorphism 
is generated by the bilinear mapping 
$B({\cal H}) \times B_1({\cal H}) \rightarrow {\bf C}$: 
$a \in B({\cal H}), b \in B_1({\cal H}) \rightarrow 
{\rm Tr}\{a \cdot b\} \in {\bf C}$, 
where ${\bf C}$ is the field of complex numbers).

For a canonical measurement of an observable $A$ on a finite-dimensional 
space state $\cal H$ (i.e., the discrete spectrum) described by Eqs. 
(\ref{CanonMap},\ref{CanonResolution},\ref{CanonRepresentation}),
the space $\Z$ coincides with the finite set of all eigenvalues 
$\lambda_i, i= 1\ldots n$ of the operator $A$ while the set $\Gamma$ 
consists of all the subsets of the set $\Z$ and for all the sets consisting
of a single point
$\{\lambda_j\}$ the operators ${\bf T}(\{\lambda_j\})$ and $M(\{\lambda_j\})$ 
are given by the formulas
\begin{equation}
   {\bf T}(\{\lambda_j\})\rho = E_j \rho E_j, \quad M(\{\lambda_j\}) = E_j.
\end{equation}
 
It is clear that the family of operators ${\bf T}(\Delta)$ provides a much 
more comprehensive description of the measurement process than the corresponding
identity resolution $M(\Delta)$  since the former allows one not only to 
calculate the statistics of obtaining various measurement outcomes, but
also determines the state of the system after the measurement is performed
(\ref{StateAfter}); generally, the same identity resolution
can be generated by different instruments ${\bf T}_1 \neq{\bf T}_2$. 

Further, it turns out \cite{Ozawa93} that for the case $\Z={\bf R}$ 
(real numbers) for any fixed $\rho$ the density matrix
$\tilde{\rho}(\Delta) = {\bf T}(\Delta)\rho$ 
allows the following integral representation: 
\begin{equation}
\label{Int_Ro}
  \tilde{\rho}(\Delta) =
  {\bf T}(\Delta)\rho=
  \int_{\Delta}\rho_\z {\rm Tr}\{\rho M(d\z)\},
\end{equation}
where $\rho_\z$ is a certain function from the space of all possible measurement
results $\Z$ into the normalized density matrices $K({\cal H})$, 
and ${\rm Tr}\{\rho M(d\z)\}$ is the ``density'' of the measure
$\mu_{\rho}$ (\ref{Prob_M})'' on $\Z$, i.e. 
\begin{eqnarray}
  \mu_{\rho}(\Delta) = {\rm Prob}(\z \in \Delta) = 
      \Tr\{\rho M(\Delta)\} = \int_{\Delta} \Tr\{\rho M(d\z)\},\\
  \mu_{\rho}(\Delta) = \int_{\Delta} d\mu_{\rho}(\z),\quad
   d\mu_{\rho}(\z) = {\rm Tr}\{\rho M(d\z)\}.
\end{eqnarray}

The function $\rho_\z$ defined in this way can already be interpreted as
the ``state of the system after the measurement which gave the outcome
$\z$''. This does not contradict to the statistical interpretation 
of quantum mechanics since actually $\rho_\z$ is only a convenient 
auxiliary tool which allows to calculate the final state of the system
after the measurement.
Physical interpretation of Eq. (\ref{Int_Ro}) is absolutely transparent
since ${\rm Tr}\{\rho M(d\z)\}$ is the probability 
of obtaining after the measurement a result in the neighborhood 
$d\z$ of point $\z$.

The reason why the representation of the type (\ref{Int_Ro}) 
is important for us is that in the teleportation algorithms
the system state after the measurement is corrected with
a unitary transformation $U_\z$ which depends on the measurement
outcome $\z$. Obviously, in this case the subensemble of systems
selected by the condition $\z \in \Delta$ after the unitary correction is 
described by the density matrix
\begin{equation}
\label{U_Ro}
\tilde{\rho}_{U,\Delta}=\int_{\Delta}U_\z\rho_\z U^+_\z\Tr\{\rho M(d\z)\};
\end{equation}
therefore, introduction of the function
$\rho_\z$ is a natural step in the attempt to extend the algorithm
of the teleportation of the state of a finite-dimensional
quantum system proposed in Ref. \cite{Bennett93} to the case of 
continuous variable. 

\section{Measurements used in quantum teleportation}
Consider now the measurements used in quantum teleportation from the viewpoint
of the general theory of quantum mechanical measurements outlines in the
preceding section. Let the particles 1 and 2 be subjected to the
measurement corresponding to the instrument
${\bf T}_{12}$. Then for the whole system including particle 3
this measurement is described by the instrument 
${\bf T}_{123}(\Delta)={\bf T}_{12}(\Delta)\otimes{\bf I}_3$, where ${\bf I}_3$
is the identity operator on $B_1({\cal H}_3)$.
Hence, after the joint measurement performed on the first and second particles
the subensemble of systems selected by the condition $\z \in \Delta$,
$\Delta \subset \Z$, $\Delta \in \Gamma$ (at this moment we do not specify
the space of possible results $\Z$) is described by the density matrix
\begin{equation}
\label{After_T12}
   \rho'_{123,\Delta} = \frac{{\bf T}_{123}(\Delta)\rho}
                    {{\rm Tr}_{1,2,3}\{{\bf T}_{123}(\Delta)\rho\}}, 
\end{equation}
and the probability of the event $\z \in \Delta$ is 
${\rm Tr}_{1,2,3}\{{\bf T}_{123}(\Delta)\rho\}$;
the reduced density matrix representing the state of particle 3 is
\begin{equation}
\label{Ro_3_After}
   \rho'_{3,\Delta} = \frac{{\rm Tr}_{1,2}\{{\bf T}_{123}(\Delta)\rho\}}
                    {{\rm Tr}_{1,2,3}\{{\bf T}_{123}(\Delta)\rho\}}. 
\end{equation}

Here we are dealing with a special case of the following more general 
situation. Suppose we have a composite system $S$ consisting of
two subsystems $A$ and $B$ whose density matrix is $\rho_{AB}$
(for the teleportation procedures the system $A$ consists of particles 
1 and 2, while the system $B$ coincides with particle 3) 
Suppose further the system $A$ is subjected a measurement described by the
instrument ${\bf T}_A$ and we wish to find the state
$\rho'_{B,\Delta}$ of the system $B$ after the measurement
(from here on the prime is used to label the state of a quantum system 
just after the measurement).
It is obvious that the instrument ${\bf T}_{AB}$ describing the change 
of the state of the entire system $A+B$ is 
${\bf T}_A\otimes {\bf I}_B$ so that 
\begin{equation}
\label{Ro_B}
   \rho'_{B,\Delta} = \frac{{\rm Tr}_{A}\{{\bf T}_{AB}(\Delta)\rho_{AB}\}}
                    {{\rm Tr}_{AB}\{{\bf T}_{AB}(\Delta)\rho_{AB}\}}. 
\end{equation}
Consider now the numerator of this fraction which, according the adopted 
conventions, will be written as
$\tilde{\rho}'_{B,\Delta}$, 
$\tilde{\rho}'_{B,\Delta} = \Tr_{A}\{{\bf T}_{AB}(\Delta)\rho_{AB}\}$ 
(so that probability of the event $\z \in \Delta$ for the measurement result $\z$ 
is ${\rm Tr}_{B}\tilde{\rho}'_{B,\Delta}$).
Let $u_B$ be an arbitrary operator from $B({\cal H})$. Let us compute
the trace ${\rm Tr}_B\{u_B\tilde{\rho}'_B\}$ (for brevity we omit everywhere 
the subscript $\Delta$):
\begin{eqnarray}
\label{Trace_u_Ro_B}
     {\rm Tr}_B\{u_B\tilde{\rho}'_B\}= \nonumber \\ 
     \Tr_B\{u_B \Tr_A\{{\bf T}_A\otimes {\bf I}_B\; \rho_{AB} \} \}= \nonumber \\
     \Tr_B\{ \Tr_A\{I_A\otimes 
               u_B\cdot{\bf T}_A\otimes {\bf I}_B\; \rho_{AB} \} \}= \nonumber \\
     \Tr_{AB}\{I_A\otimes u_B\cdot{\bf T}_A\otimes 
              {\bf I}_B\; \rho_{AB} \}= \nonumber \\
     \Tr_{AB}\{\left[({\bf T}_A\otimes {\bf I}_B)^* I_A\otimes u_B\right]
                  \cdot \rho_{AB} \} = \nonumber \\
     \Tr_{AB}\{\left[({\bf T}_A^* I_A)\otimes {\bf I}_B^* u_B\right]
                  \cdot \rho_{AB} \} = \nonumber \\
     \Tr_{AB}\{\left[M_A\otimes u_B\right]
                  \cdot \rho_{AB} \} = \nonumber \\
     \Tr_{AB}\{\left[(M_A\otimes I_B)\cdot (I_A\otimes u_B)\right]
                  \cdot \rho_{AB} \} = \nonumber \\
     \Tr_{AB}\{\left[(I_A\otimes u_B)\cdot (M_A\otimes I_B)\right]
                  \cdot \rho_{AB} \} = \nonumber \\
     \Tr_{B}\{ \Tr_A\{\left[(I_A\otimes u_B)\cdot (M_A\otimes I_B)\right]
                  \cdot \rho_{AB} \} \}= \nonumber \\
     \Tr_{B}\{ u_B\Tr_A\{(M_A\otimes I_B) \cdot \rho_{AB} \} \}.
\end{eqnarray}
Therefore,
\begin{equation}
    \tilde{\rho}'_{B,\Delta} = \Tr_{A}\{{\bf T}_{AB}(\Delta)\rho_{AB}\} = 
       \Tr_A\{(M_A(\Delta)\otimes I_B) \cdot \rho_{AB} \}.
\end{equation}
Thus, if we wish to find the state of the system $B$ just after the measurement
performed over the system $A$, it is sufficient for us to know
only the identity resolution in ${\cal H}_A$ on $\Z$ generated by the 
instrument ${\bf T}_A$ rather than the instrument ${\bf T}_A$ itself.

It should be noted that the technique of quantum operations
seems first to have been applied to the problem of teleportation in the work
\cite{NielsenCaves96} where the simplest case of ``ideal'' teleportation
with the discrete space of possible measurement results $\Z$ was considered
when the change of the system state caused by the measurement is described by 
the instrument of the type
\begin{equation}
\label{Ideal}
  \rho \rightarrow A_i \rho A_i^+, 
\end{equation}
where $A_i$ is a positive operator and the subscript $i=1,2\ldots$ 
labels different measurement results, i.e., points of $\Z$. 
However, in that work the teleported state was expressed through the
operators $A_i$ which completely characterize the entire instrument.

We are interested in the possibility of the
representation of $\tilde{\rho}'_{B,\Delta}$ in the form
\begin{equation}
 \label{Int_dmu}
 \tilde{\rho}'_{B,\Delta} = \int_\Delta  \rho_{\z,B} d\mu_{{\rho}_{AB}}(\z),
\end{equation}
where $\rho_{\z,B} \in K({\cal H}_B)$, 
and the measure $d\mu_{\rho_{AB}}(\z)$ is the probability density 
of obtaining measurement result in the neigbourhood of point $\z$, 
i.e. satisfies the condition
\begin{equation}
{\rm Tr}_{B}\tilde{\rho}'_{B,\Delta} = \int_\Delta d\mu_{{\rho}_{AB}}(\z).
\end{equation}
Formally, such a representation can be easily found if 
the measure $\mu_{{\rho}_{AB}}$ is absolutely continuous 
with respect to the initial measure $d\z$ on $\Z$ and the matrix elements
of the operators $M_A(\Delta)$ calculated for some orthogonal basis
$\ket{\varphi_{nA}}$ of the system $A$ can be written as
\begin{equation}
\label{Matrix_IntRep}
  \bra{\varphi_{mA}} M_A(\Delta) \ket{\varphi_{nA}} = 
           \int_\Delta d\z F_{mn}(\z),
\end{equation}
where $F_{mn}(\z)$ are the $c$-numbers valued functions on $\Z$
(for example, if the measurement $M$ corresponds to a simultaneous 
measurement of the complete set of commuting observables with the continuous
spectrum, since in that case
${\cal H}_A=L^2(\Z)$, while the space $\Z$ is a direct product of
the spectra of the operators comprising this set so that
$F_{mn}(\z)=\varphi_{mA}(\z)^*\psi_{nA}(\z)$).
Indeed, in this case
\begin{eqnarray}
\label{Int_dx}
    \tilde{\rho}'_{B,\Delta} = 
       \Tr_A\{(M_A(\Delta)\otimes I_B) \cdot \rho_{AB} \} 
        = \sum_{mn}
  \bra{\varphi_{mA}} M_A(\Delta) \ket{\varphi_{nA}}  \rho_{nm,B}\nonumber \\
        = \sum_{mn}
           \int_\Delta d\z\; F_{mn}(\z)\; \rho_{nm,B} 
        =
           \int_\Delta d\z \left[\sum_{mn} F_{mn}(\z)\; \rho_{nm,B}\right] = 
           \int_\Delta d\z\; \tilde{\rho}_{\z,B},
\end{eqnarray}
where the operator $\rho_{nm,B}$ on ${\cal H}_B$ is obtained from the
operator $\rho_{AB}$ by taking a ``partial matrix element'' over the vectors
$\varphi_{nA}$ and $\varphi_{mA}$ from ${\cal H}_A$,
\begin{equation}
  \rho_{nm,B} = 
  \bra{\varphi_{nA}} \rho_{AB} \ket{\varphi_{mA}},  
\end{equation}
and
\begin{equation}
    \tilde{\rho}_{\z,B} = \sum_{mn} F_{mn}(\z)\; \rho_{nm,B}.
\end{equation}
Hence 
\begin{equation}
   {\rm Tr}_{B}\{\tilde{\rho}'_{B,\Delta}\} = 
     \int_\Delta d\mu_{{\rho}_{AB}}(\z)=
     \int_\Delta d\z \Tr_B\{\tilde{\rho}_{\z,B}\}. 
\end{equation}
Therefore, multiplying and dividing the integrand in the last integral
in Eq. (\ref{Int_dx}) by $H(\z) = \Tr\{\tilde{\rho}_{\z,B}\} > 0$,
we obtain Eq. (\ref{Int_dmu}) where
\begin{equation}
   \rho_{\z,B} = \frac{\tilde{\rho}_{\z,B}} 
                     {\Tr\{\tilde{\rho}_{\z,B}\}}=
                \frac{\tilde{\rho}_{\z,B}} 
                     {H(\z)},
\end{equation}
so that $\Tr\{\rho_{\z,B}\} = 1$ and $d\mu\rho_{AB}(\z) = H(\z)d\z$,
i.e. $H(\z)$ is the Radon-Nikodim derivative of the measure 
$d\mu_{{\rho}_{AB}}(\z)$ with respect to measure $d\z$. 
We shall not dwell on the correctness of the procedure of changing the order
of summation of an infinite series and integration in Eq. (\ref{Int_dx}) 
and other similar operations since in the particular cases
considered in the rest of the paper the integral representation
of the form (\ref{Int_dx}) directly follows from the specific
from of the operators $M(\Delta)$.

\section{Teleportation with an orthogonal measurement}
To illustrate the outlined general scheme, we shall first consider
the teleportation of an unknown quantum state $\ket{\psi}$ 
of a one-dimensional non-relativistic spinless particle.
To avoid the complications associated with the particle permutation 
symmetry we shall assume that all three particles are different.
It is sufficient to consider the case where the initial state of particle 
1 is a pure state
\begin{equation}
\rho_1 = \rho_\psi = \ket{\psi;1}\bra{\psi;1},  \quad
\ket{\psi; 1} = \int_{-\infty}^{+\infty} dx \psi(x) \ket{x; 1}.
\end{equation}
The entangled state of particles 2 and 3 will be chosen in the form
of an EPR-state (with an infinite norm)
\begin{equation}
\label{EPR_state}
\rho_{23} = \ket{\psi_{23}}\bra{\psi_{23}}, \quad
\ket{\psi_{23}}=
\int_{-\infty}^{\infty}dx \ket{x;2} \ket{x;3}, 
\end{equation} 
which can be represented as a limit of a normalized state
\begin{equation}
\ket{\Psi_{23}}=
\int_{-\infty}^{\infty}
\int_{-\infty}^{\infty} dx dy
\Psi(x,y)
\ket{x;2}\ket{y;3},
\end{equation} 
where 
$\Psi(x,y)\rightarrow \delta(x-y)$ (in the momentum representation
$\Psi_{23}(p_1,p_2) \rightarrow \delta(p_1+p_2)$);
formally, the state (\ref{EPR_state}) is an eigenvector of
the operator of the difference of the positions of the second 
and the third particle: 
$(X_2 - X_3)\ket{\psi_{23}}=0$.

Consider now the joint measurement performed over one of the particles from
the EPR-pair (particle 2) and the system in the unknown state to be teleported
(particle 1) defined by the following identity resolution:
\begin{equation}
  \int_{-\infty}^{\infty}\int_{-\infty}^{\infty}
  E_{12}(dXdP)=I,
\end{equation} 
\begin{equation}
\label{dXdPResolution}
E_{12}(dXdP)= \KetBra{\Phi_{XP}}{\Phi_{XP}} \frac{dXdP}{2\pi} =
\end{equation} 
\begin{equation}
 =\frac{1}{2\pi}
  \int_{-\infty}^{\infty}dx  
  \int_{-\infty}^{\infty}dx' 
  e^{iP(x-x')} \ket{x+X;1}  \ket{x;2} 
              \bra{x'+X;1} \bra{x';2}  
 dXdP,
\end{equation}
where
\begin{equation}
\label{EigenXP}
  \ket{\Phi_{XP}} = \int_{-\infty}^{\infty}dx e^{iPx} \ket{x+X;1} \ket{x;2}; 
\end{equation}
note that formally the state (\ref{EigenXP}) is a common eigenvector
for a pair of the commuting observables $X_1 - X_2$ and $P_1 + P_2$
(position difference and the total momentum) which form a complete set
of commuting observables on the state space of two particles:
$(X_1 - X_2) \ket{\Phi_{XP}}= X \ket{\Phi_{XP}}$,
$(P_1 + P_2) \ket{\Phi_{XP}}= P \ket{\Phi_{XP}}$;
therefore, the teleportation procedure with $\rho_{23}$ taken in the form 
(\ref{EPR_state}) and the measurement
(\ref{dXdPResolution}) is exactly coincides with the algorithm 
\cite{Vaidman94}. In that case
the space of all possible measurement results $\Z$ is the set of ordered
pairs $(X,P)$, $(-\infty<X<\infty,\; -\infty<P<\infty)$ constituting a plane
${\bf R}^2$ which is actually a direct product of two copies of the real line
${\bf R}_X$ and ${\bf R}_P$  corresponding to the position  $X$ and momentum 
$P$: $\Z = {\bf R}_X \times {\bf R}_P$.
 
The exact meaning of Eq. (\ref{dXdPResolution}) is that the
matrix elements of the positive operator $E(\Delta)$ associated with the set
$\Delta$ can be calculated as
\begin{equation}
  \bra{\Phi} E_{12}(\Delta) \ket{\Psi} = 
  \int_{\Delta}\frac{dXdP}{2\pi} 
  \int_{-\infty}^{\infty}dx  
  \int_{-\infty}^{\infty}dx' 
  e^{iP(x-x')} \Phi^*(x+X,x) \Psi(x'+X,x'),
\end{equation}
similar to Eq. (\ref{Matrix_IntRep}).

Simple calculations reveal that the teleported density matrix in channel 3 
becomes
\begin{equation}
\label{Ro_3}
  \tilde{\rho}'_{3,\Delta} =
  \Tr_{1,2}\{(\rho_1\otimes\rho_{23})E_{12}(\Delta)\} =
  \int_{\Delta}\,\rho_{XP}\,\, \frac{dXdP}{2\pi},
\end{equation}
where
\begin{equation}
\label{psi_3}
  \rho_{XP}=
  \ket{\psi_{XP};3}\bra{\psi_{XP};3},\quad
  \psi_{XP}(x) = e^{iPx} \psi(x+X).
\end{equation}
Since 
\begin{equation}
  \Tr_3\{\rho_{XP}\} = \int_{-\infty}^{+\infty}dx\left|\psi(x+X)\right|^2 = 1, 
\end{equation}
it is clear that the probability density of obtaining after the measurement
a result in the interval $(dX,dP)$ in the neighbourhood of point $(X,P)$ 
is $1/2\pi$ and does not depend on $\ket{\psi;1}$,
so that the measurement does not provide any information on the
teleported state. Total probability of obtaining any pair $(X,P)$ turns out to 
be infinite because of the unnormalizabilty of state (\ref{EPR_state}).

Eqs. (\ref{Ro_3}-\ref{psi_3}) imply that subjecting the particle 3 to
a unitary transformation
\begin{equation}
  U_{XP}:\;\; \psi(x) \rightarrow e^{iP(x-X)}\psi(x-X)
\end{equation}
which only depends on the result of the measurement performed over particles
1 and 2, one obtains in the channel 3 the state identical to the initial state
of particle 1, i.e. achieves the teleportation of the state of particle 1.
It should be noted that in the present example the unitary correction
(which does not depend on $\rho_1$) of the state of the third particle to
the initial state of particle 1 proves to be possible for any input
state $\rho_1$ and any measurement outcome, i.e. any pair $(X,P)$.
However, it is generally reasonable to consider also the teleportation 
algorithms which allow the teleportation of only a subset $K'({\cal H}_1)$
of all possible states, e.g. only the states belonging to a certain
subspace ${\cal H}'_1 \subset {\cal H}_1$ rather than the total space
${\cal H}_1$ \cite{NielsenCaves96} 
(an example of that kind of algorithm is presented in the next section).
In addition, the requirement that the necessary unitary correction $U_\z$
exists for all measurement outcomes is also unnecessary.
Indeed, the entire space of all possible measurement outcomes
$\Z$ can always be divided into two disjoint subsets $\Z_1$ and $\Z_2$,
$\Z= \Z_1 \cap \Z_2 = \emptyset$, $\Z = \Z_1 \cup \Z_2$, in the following way:
an arbitrary point $\z \in \Z$ belongs to the subset $\Z_1$ if and only if
the unitary transformation $U_\z$ with the required properties 
exists. A sufficient condition for the possibility of the teleportation
will than be a non-zero measure
$\mu_\rho(\Z_1)$ for all $\rho \in K'({\cal H}_1)$.
In that case the teleportation algorithm looks as follows:
the ensemble of systems representing the initial state $\rho_1$ is subjected
to the joint (together with the particle 2) to the measurement
$m_{12}$. If the outcome 
$\z \in \Z_2$, then the particular copy of the system 3 is discarded. 
On the other hand, if $\z \in \Z_1$,
then the system 3 is subjected to the unitary correction $U_\z$. 
Under these conditions the subensemble of particles 3 selected and corrected 
in the above outlined way will be found in the state identical to the
initial state $\rho_1$ of particle 1.

\section{Teleportation with a non-orthogonal measurement}
We shall now consider an example of the teleportation of an unknown state
based on the measurement associated with a non-orthogonal identity resolution.
Consider a model quantum system whose Hamiltonian has a continuous 
non-degenerate spectrum coinciding with the positive part of the real line
$(0,+\infty)$ (e.g. a free non-relativistic one-dimensional spinless
particle whose allowed states are restricted by the condition of only positive 
momentum components occurring in their momentum representation).
Thus we shall assume that an arbitrary pure state of the system 1 can
be described by a wave function defined on the positive part of the real line:
\begin{equation}
 \ket{\psi;1} = \int_0^{+\infty} \psi(E)\ket{E;1}dE, \quad 
 \BraKet{E}{E'}=\delta{(E-E')}.
\end{equation}
The EPR-state in the energy representation can be chosen, e.g. in the form
\begin{equation}
\label{EPR_omegat}
\ket{\psi_{23}}=
\int_{0}^{\ve_0} 
d \ve
\ket{\ve;2}
\ket{\ve_0-\ve;3}.
\end{equation}
Such an EPR-pair can be considered as the limit of a normalized state
\begin{equation}
\ket{\Psi_{23}}=
\int_{0}^{\ve_0} \int_{0}^{\ve_0} 
d \ve_1 d \ve_2
\psi(\ve_1,\ve_2)
\ket{\ve_1;1}
\ket{\ve_2;2},
\end{equation}
where 
$\psi(\ve_1,\ve_2)\rightarrow
\delta(\ve_1+\ve_2-\ve_0)$. 
The states of that kind are produced in the parametric down-conversion
if the pump frequency is $\ve_0$. 
Formally, the EPR-state can also be chosen in the form 
$\psi(\ve_1,\ve_2)\rightarrow \delta(\ve_1-\ve_2)$;
however, it is not clear how this state can be realized experimentally.

Consider now a joint measurement $M_{12}(d\Omega dT)$ on the particles 
1 and 2 defined by the following non-orthogonal identity resolution
\begin{displaymath}
  M_{12}(d\Omega dT)=
\end{displaymath}
\begin{eqnarray}
  \label{dOmega_dT}
  =\frac{1}{\pi}
   \left(
     \int_{-\Omega}^{\Omega} d\omega e^{i\omega T}
     \ket{\Omega + \omega; 1}\ket{\Omega - \omega; 2} 
   \right)
   \left(
     \int_{-\Omega}^{\Omega} d\omega' e^{-i\omega' T}
     \bra{\Omega + \omega'; 1}\bra{\Omega - \omega'; 2} 
   \right) d\Omega dT \\
   =\frac{1}{\pi}
   \int_{-\Omega}^{\Omega}  
   \int_{-\Omega}^{\Omega} d\omega d\omega' e^{i(\omega - \omega') T}
   \ket{\Omega + \omega; 1}\ket{\Omega - \omega; 2} 
   \bra{\Omega + \omega'; 1}\bra{\Omega - \omega'; 2} d\Omega dT; 
\end{eqnarray}
here $\Omega$ and $T$ vary in the ranges 
${\bf R}_{\Omega}^+ = (0;+\infty)$ and ${\bf R}_T = (-\infty; +\infty)$, 
respectively, so that the space of all possible measurement results
is $\Z={\bf R}_{\Omega}^+ \times {\bf R}_T$.
The quantities $\Omega$ and $\omega$ have the meaning of the
half-sum and half-difference of the energies of two particles
(we do not distinguish energy and frequency), 
e.g., two photons in the biphoton.
This measurement which in some sense is an intermediate measurement between 
the frequency and time parameter measurement for two-particle states
can be realized, at least in principle, for the photons experimentally 
employing the parametric up-conversion phenomenon \cite{Molotkov_2}. 

It is easily checked that $M_{12}(d\Omega dT)$ is actually 
an identity resolution:
\begin{displaymath}
  \int_{0}^{\infty}\int_{-\infty}^{\infty}M_{12}(d\Omega dT)=
\end{displaymath}
\begin{displaymath}
  \frac{1}{\pi}
  \int_{0}^{\infty}d\Omega
  \int_{-\infty}^{+\infty} dT
  \int_{-\Omega}^{\Omega} d\omega
  \int_{-\Omega}^{\Omega} d\omega'
  e^{i(\omega-\omega')T}
  \ket{\Omega + \omega; 1}\ket{\Omega - \omega; 2} 
  \bra{\Omega + \omega'; 1}\bra{\Omega - \omega'; 2} = 
\end{displaymath}
\begin{displaymath}
  2 \int_{0}^{\infty}d\Omega
  \int_{-\Omega}^{\Omega} d\omega
  \int_{-\Omega}^{\Omega} d\omega'
  \delta(\omega-\omega')
  \ket{\Omega + \omega; 1}\ket{\Omega - \omega; 2} 
  \bra{\Omega + \omega'; 1}\bra{\Omega - \omega'; 2} = 
\end{displaymath}
\begin{displaymath}
  \int_{0}^{+\infty} d\omega_1
  \int_{0}^{+\infty} d\omega_2
  \ket{\omega_1; 1}\ket{\omega_2; 2} 
  \bra{\omega_1; 1}\bra{\omega_2; 2} =  I_{12},
\end{displaymath}
where $\omega_1 = \Omega + \omega$ and $\omega_2 = \Omega - \omega$. 

The teleported density matrix is now 
\begin{equation}
\label{Ro_3_OmegaT}
  \tilde{\rho}'_{3,\Delta} =
  \Tr_{1,2}\{(\rho_1\otimes\rho_{23})M_{12}(\Delta)\} =
  \int_{\Delta}\,\rho_{\Omega T}\,\, \frac{d\Omega dT}{\pi}, \quad
  \rho_{\Omega T}=
  \ket{\psi_{\Omega T};3}\bra{\psi_{\Omega T};3},
\end{equation}
where (for brevity we write $\ket{\psi_3}$ instead of $\ket{\psi_{\Omega T};3}$) 
\begin{equation}
\label{psi_3_OmegaT}
  \ket{\psi_{\Omega T};3} = 
  \ket{\psi_3} = \int_{\ve_0 - {\rm min}\{\ve_0,2\Omega\}}^{\ve_0}
  d\ve e^{-i(2\Omega - \ve_0 + \ve)T}
  \psi(2\Omega - \ve_0 + \ve) \ket{\ve;3}.
\end{equation}
The probability of obtaining the measurement result in the interval
$(\Omega,\Omega+d\Omega; T,T+dT)$ is
\begin{equation}
\label{Prob_OmegaT}
  \Tr\{\tilde{\rho}'_{d\Omega dT}\}=
  \Tr_{1,2,3}\{(\rho_1\otimes\rho_{23})M_{12}(d\Omega dT)\}=
  \frac{d\Omega dT}{\pi}
  \int_{\ve_0 - {\rm min}\{\ve_0,2\Omega\}}^{\ve_0}
  |\psi(2\Omega - \ve_0 + \ve)|^2 d\ve.
\end{equation}
Note that the corresponding probability density 
does not depend on $T$.
Since $T$ varies in the infinite interval, the total probability,
just as in the preceding section, proves to be infinite. 
Formally, this is related to the fact that the 
state (\ref{EPR_omegat}) has an infinite norm. However, this 
circumstance does not create any problems since all the physically 
meaningful results can be calculated on the basis of the relative 
probabilities of different events.

Suppose now that the support of wave function $\psi$ of the system 1
is known to lie in a certain segment $[E_{\rm min},E_{\rm max}]$,
i.e. $\psi(E) = 0$ at $E>E_{\rm max}$ and $E<E_{\rm min}$. 
In that case the probability density (\ref{Prob_OmegaT}) does depend
on $\Omega$; e.g. it vanishes for 
$2\Omega > E_{\rm max} + \ve_0$, since then the function $\psi$ 
is zero in the entire integration interval.
Clearly, the condition for the exact teleportation is that
the support of function $\psi$ should belong to the integration
interval in Eq.(\ref{psi_3_OmegaT}); in that case the probability of 
obtaining a particular result $\Omega$ does not depend on $\ket{\psi;1}$
since the integral in Eq. (\ref{Prob_OmegaT}) is identically equal to 1
due to the normalization of $\ket{\psi; 1}$.

It is convenient to perform the further analysis for the cases 
$\ve_0 > E_{\rm max}$ and $\ve_0 < E_{\rm max}$ separately. 
Consider first the case $\ve_0 > E_{\rm max}$. 
If the measurement gave the result $2\Omega < \ve_0$ (case 1a),  
the state of the system 3 will be $\ket{\psi_3}\bra{\psi_3}$, where
\begin{equation}
\label{first_case}
 \ket{\psi_3} = \int_\gamma^{\ve_0}
  d\ve e^{-i(\ve-\gamma)T} 
  \psi(\ve - \gamma)\ket{\ve;3}, \quad
  \gamma = \ve_0 - 2\Omega.
\end{equation}
Here the argument of the function $\psi$ in the integrand ranges 
from 0 to $2\Omega$. Therefore, 
the state $\psi$ can only be teleported if its support 
$[E_{\rm min}, E_{\rm max}] \subset [0,2\Omega]$, i.e. if 
$E_{\rm max}< 2\Omega$.
Thus, $[E_{\rm max}, \ve_0] \subset \Z_1$
(we omit the trivial factor ${\bf R}_T$ in $\Z_1$,
since the value of $T$ does not matter). 

On the other hand, if the measurement gave the result 
$2\Omega > \ve_0$ (case 1b), the 
state of the system 3 will be $\ket{\psi_3}\bra{\psi_3}$, where
\begin{equation}
\label{second_case}
 \ket{\psi_3} = \int_0^{\ve_0}
  d\ve e^{-i(\ve+\gamma)T} 
  \psi(\ve + \gamma)\ket{\ve;3}, \quad
  \gamma = 2\Omega - \ve_0.
\end{equation}
Now the argument of the function $\psi$ in the integrand ranges
from $\gamma$ to $2\Omega$ and 
the state $\psi$ can only be teleported if its support 
$[E_{\rm min}, E_{\rm max}] \subset [\gamma,2\Omega]$, i.e. if 
$\gamma< E_{\rm min}$ or, in other words, $2\Omega < \ve_0 + E_{\rm min}$ 
(the condition $E_{\rm max}< 2\Omega$ is fulfilled automatically since
$2\Omega> \ve_0 > E_{\rm max}$). 
Thus, $[\ve_0, \ve_0 + E_{\rm min}] \subset \Z_1$.
Bringing the case 1a and 1b together we obtain 
$\Z_1 = [E_{\rm max}, \ve_0 + E_{\rm min}]$.

It is seen from Eqs. (\ref{first_case}) and (\ref{second_case}) 
that the state of the system 3 can be made identical 
to the initial state of the system 1 just before the measurement if the system
3 just after the measurement is subjected to the unitary transformation
\begin{equation}
\label{Unitar_1}
\psi(\ve)\rightarrow \tilde{\psi}(\ve) = 
  \left\{ \begin{array}{ll}
            \psi(\ve),                     & \mbox{ if\ \ } \ve > \ve_0  \\
            \psi(\ve + \gamma) e^{i\ve T}, & \mbox{ if\ \ } 0 < \ve < 2\Omega\\
            \psi(\ve - 2\Omega),           & \mbox{ if\ \ } 2\Omega < \ve < \ve_0
          \end{array}
  \right.
\end{equation}
or
\begin{equation}
\label{Unitar_2}
\psi(\ve)\rightarrow \tilde{\psi}(\ve) = 
  \left\{ \begin{array}{ll}
            \psi(\ve),                     & \mbox{ if\ \ } \ve > 2\Omega\\
            \psi(\ve + \gamma) e^{i\ve T}, & \mbox{ if\ \ } \gamma < \ve < 2\Omega\\
            \psi(\ve - 2\Omega),           & \mbox{ if\ \ } 0 < \ve < \gamma
          \end{array}
  \right.
\end{equation}
in the cases 1a and 1b, respectively. 

Consider now the situation where $\ve_0 < E_{\rm max}$. In that case
the measurements yielding the results $2\Omega < \ve_0$ should certainly
be discarded since the range of the variation of the argument of function 
$\psi$ in Eq. (\ref{first_case}) does not cover the support of function $\psi$.
However, if the measurement gave the result $2\Omega > \ve_0$ then, similar to
the case 1b, the teleportation is still possible (using the unitary 
transformation (\ref{second_case}), if
$[E_{\rm min}, E_{\rm max}] \subset [\gamma,2\Omega]$, i.e. if 
the conditions $\gamma< E_{\rm min}$ 
(i.e. $2\Omega < \ve_0 + E_{\rm min}$) and
$E_{\rm max}< 2\Omega$ are simultaneously satisfied (the latter inequality
now imposes an additional constraint rather than being satisfied 
automatically). The existence of an interval of the values of
$\Omega$ where the inequalities
$2\Omega < \ve_0 + E_{\rm min}$ and $E_{\rm max}< 2\Omega$ are simultaneously
satisfied is only possible if the inequality 
$E_{\rm max}< E_{\rm min} + \ve_0$ hold, or, in other words
if $\ve_0 > E_{\rm max} - E_{\rm min}$. 
In that case again $\Z_1 = [E_{\rm max}, \ve_0 + E_{\rm min}]$.
Thus in the proposed scheme the teleportation is possible if and only if
the spectrum width of the EPR-pair (\ref{EPR_omegat}) exceeds the spectral width of the support 
of function $\psi$. 

It should be noted the teleportation of a broadband single-photon wave packet
was first considered in \cite{Molotkov_1,Molotkov_2}. Recently,
the algorithm for the teleportation of a single-mode electromagnetic field
based on the squeezed states \cite{BraunKimble98} was extended to the
case of a broadband input state \cite{LoockBraunKimble} whose 
spectral density is restricted to the vicinity of the half-frequency
of the pump field producing the indicated squeezed state.
In contrast to the algorithm proposed in the present paper,
the scheme of Ref.\cite{LoockBraunKimble} employs the orthogonal measurements.
Physically, the non-orthogonal measurement (\ref{dOmega_dT}) 
is naturally arising when considering the system states in the energy
representation: just as the originally proposed teleportation scheme
formulated in the position representation \cite{Vaidman94} actually employs
the simultaneous measurement of position and momentum, it is natural to 
suppose that a similar procedure can be implemented measuring
the energy and the conjugated quantity, i.e. time. However, since
in quantum mechanics the time observable is not associated with any
self-adjoint operator, the resulting measurement turns out to be 
non-orthogonal (and, of course, the involved EPR-pair is entangled in energy 
rather than position).

It should also be noted that Ref.\cite{BraunKimble98} addressed 
the teleportation of a quantum state described by dynamic variables 
$(x,p)$ (the unknown state in Ref.\cite{BraunKimble98} 
corresponds to the single-mode photon state) for the case 
of a non-ideal EPR-pair (squeezed state). Non-ideality of the 
EPR-correlation reduces the accuracy (fidelity) of the teleportation.
The example based on an orthogonal measurement shows that 
the singular EPR-states allows the achievement of an unconditional
exact $(\rm{fidelity} = 1)$ teleportation.  The word ``unconditional''
here means that any measurement outcome leads to an exact teleportation.
In the case of the proposed non-orthogonal measurement
the unconditional exact teleportation is impossible even with a singular
EPR-pair since for some measurement outcomes there exist no unitary 
transformations recovering the exact copy of the initial input state
in the channel 3; these outcomes should be discarded. All the left
outcomes provide an exact teleportation.

The experiments on teleportation may involve the situation
when actually realized instead of a theoretically unconditional
measurement (leading to the exact teleportation for any outcome) is
its certain approximation and the teleportation becomes conditional 
even if one assumes that the experiment uses an ideal EPR-pair.
Formally, any measurement is described by an identity resolution;
the experimental implementation of a particular identity resolution
requires the selection of a suitable interaction between the quantum system
and the measuring device reproducing the necessary space of all possible
measurement outcomes an the probability density distribution on that space
specified by the given identity resolution. Usually this a very difficult task
even for the systems with the discrete degree of freedom (e.g. spin or
polarization). Therefore, the surplus outcomes arise which should be discarded.
For example, the non-orthogonal identity resolution
(\ref{dOmega_dT}) can be realized through the coalescence of
a pair of photons in a non-linear crystal (parametric up-conversion)
and subsequent detection of the arising photon \cite{Molotkov_2}.
However, because of the small non-linear perceptibility,
a lot of idle outcomes (when the photodetector does not fire) occur
which should be discarded.

\section{Conclusions}
In summary, we have analyzed the measurements used in quantum teleportation
from the viewpoint of the general quantum mechanical theory of
measurements. It is shown that the teleported state is completely determined
by the identity resolution (positive operator-valued measure) 
in the system state space generated by the corresponding instrument
(quantum operation describing the change of the system state caused by
the act of measurement) rather than the instrument itself, so that it is
not actually necessary to specify the instrument itself providing
the most complete description of the measuring procedure allowed by
general laws of quantum mechanics. 
An algorithm for the teleportation of the state of a quantum
system with a continuous non-degenerate spectrum is proposed based
on a non-orthogonal measurement. Similar to all other available
protocols providing an exact teleportation, our protocol
employs an ideal EPR-pair with the singular correlations
which corresponds to an unnormalizable wave function\footnote{Strictly 
speaking, the correct analysis of such states 
should be based on the rigged Hilbert state approach; in our case 
the problem of the infinite norm of employed  
EPR-states is avoided since we are only interested in
the relative probabilities of different events}. 
It should be noted that the question of the possibility 
of achieving an exact teleportation of continuous quantum variable
with the physically realizable (normalized) states is still open;
for example, no algorithms of exact quantum teleportation for
continuous variable have yet been proposed with non-singular EPR-states.

The authors are grateful to Prof. K.A.Valiev for the discussion of 
obtained results. The work was supported by the Russian Fond for Basic 
Research (project No 99-02-18127) and by the Program ``Advanced
technologies and devices of nano- and microelectronics''
(project No 02.04.5.2.40.T.50).


\begin{thebibliography}{99}
\bibitem{Bennett93} C.H.Bennett, G.Brassard, C.Crepeau, R.Jozsa, A.Peres, and
   W.K.Wootters, Phys. Rev. Lett., {\bf 70}, 1895 (1993).
\bibitem{Vaidman94} L.Vaidman, Phys. Rev., {\bf A49}, 1473 (1994).
\bibitem{BraunKimble98} S.Braunstein, H.J.Kimble, Phys. Rev. Lett., 
   {\bf 80}, 869 (1998).
\bibitem{Ozawa93} M.Ozawa, J. Math. Phys., {\bf 34}, 5596 (1993). 
\bibitem{Kraus} K.Kraus, {\it States, Effects and Operations}, Springer-Verlag, 
   Berlin, 1983.
\bibitem{Holevo} A.S.Holevo, {\it Probabilistic and Statistical Aspects of
   Quantum Theory}. North Holland Publishing Corporation, Amsterdam, 1982.
\bibitem{Neumann} J. von Neumann, {\it Mathematical Foundations of Quantum 
   Mechanics}, Princeton Uni\-ver\-sity, Princeton, NJ, 1955.
\bibitem{Luders} G.L\"uders, Ann. Physik, {\bf 8} (6), 322 (1951).
\bibitem{NielsenCaves96} M.A.Nielsen and C.M.Caves, 
      E-Print http://xxx.lanl.gov/quant-ph/9608001.
\bibitem{Molotkov_1} S.N.Molotkov, Phys. Lett., A {\bf 245}, 339 (1998)\\
   and E-Print http://xxx.lanl.gov/quant-ph/9805045.
\bibitem{Molotkov_2} S.N.Molotkov, ZhETF Letters, {\bf 68}, 248 (1998)\\
   and E-Print http://xxx.lanl.gov/quant-ph/9807013.
\bibitem{LoockBraunKimble} P. van Loock, S.Braunstein, and H.J.Kimble, 
   E-Print http://xxx.lanl.gov/quant-ph/9902030.

\end{thebibliography}
\end{document}